\documentclass[aps,pra,twocolumn,superscriptaddress,showpacs,nofootinbib,floatfix]{revtex4-2}
\usepackage{amsmath,amssymb}
\usepackage{graphicx}
\usepackage{dcolumn}
\usepackage{bm}
\usepackage[normalem]{ulem}
\newcolumntype{d}{D{.}{.}{-1}}
\usepackage[colorlinks=true,linkcolor=blue,citecolor=blue,urlcolor=blue]{hyperref}

\newcommand{\Rr}{\mathcal{R}}
\newcommand{\vpar}{v_\parallel}
\newcommand{\vperp}{v_\perp}
\newcommand{\Eperp}{E_\perp}
\newcommand{\bq}{\bm{q}}
\newcommand{\bK}{\bm{K}}
\newcommand{\br}{\bar{\bm{r}}}
\newcommand{\bX}{\bm{X}}
\newcommand{\bE}{\bm{E}}
\newcommand{\hz}{\hat{\bm{z}}}
\newcommand{\hq}{\hat{\bm{q}}}
\newcommand{\dd}{\mathrm{d}}

\begin{document}

\title{Surface charges and quantum reflection of hydrogen-isotope beams at grazing incidence}

\author{A.\ Yu.\ Voronin}
\affiliation{P.\,N.\ Lebedev Physical Institute, 53 Leninsky Prospect, 119991 Moscow, Russia}
\author{V.\ V.\ Nesvizhevsky}
\affiliation{Institut Max von Laue -- Paul Langevin, 71 avenue des Martyrs, F-38042 Grenoble, France}

\date{\today}

\begin{abstract}
Experiments with cryogenic beams of hydrogen isotopes (hydrogen (H), deuterium (D), and tritium (T)) can be used to study quantum reflection from solid dielectric mirrors. Most dielectric surfaces carry trapped electric charges whose field polarizes the incident atom. We study theoretically how such charges, distributed randomly, in clusters, with compensation over a finite length, or in order, change specular reflection and diffuse scattering of H, D, and T at grazing incidence on typical dielectric mirrors (SiO$_2$/Si and MgF$_2$), for longitudinal velocities of 100--300~m/s and normal velocities of 3--100~cm/s which might become available soon in grazing incidence experiments. Starting from the exact coupled-channel equations, we show that the specular wave feels only the laterally averaged polarization potential, which is fixed by the structure factor of the charges and, for uncorrelated charges, is an inverse-square tail $-C_2/z^2$. The exact solution replaces the linear threshold law $1-\Rr\propto\vperp$ by the fractional power $\vperp^{2\nu}$, $\nu=(1/4-\zeta)^{1/2}$, with the coupling $\zeta\propto C_2$; for weak coupling, this appears as a logarithmic velocity dependence, and for $\zeta>1/4$, the reflection coefficient oscillates log-periodically. At the surface charge density of $10^{12}$ cm$^{-2}$, absorption increases by up to 26\% for SiO$_2$/Si and by a factor of two to three for MgF$_2$; the effect increases with atomic mass, saturates for compensated charges, and is accompanied by diffuse scattering below $10^{-3}$ of charge-induced loss. We also formulate the condition of complete absorption at the solid surface. 

\end{abstract}

\maketitle

\section{Introduction}
\label{sec:intro}

A slow atom approaching a surface is reflected by the attractive atom--surface interaction potential without reaching the classical turning point \cite{Friedrich2002,Friedrich2004,Cote1998,Jurisch2006,Madronero2007}. Quantum reflection (QR) of hydrogen (H) atoms was observed from the surface of liquid helium \cite{Berkhout1989,Yu1993}, metastable neon and helium atoms, as well as helium atoms and sodium condensates from solid surfaces \cite{Shimizu2001,Druzhinina2003,Pasquini2004,Oberst2005,ShimizuFujita2002,Oberst2005PRL,Pasquini2006,Zhao2008,Zhao2011,Bender2010,Garcion2021,Lecoffre2025,Kilianski2024}. It is now studied experimentally for the lightest atoms, H and its isotopes (deuterium (D) and tritium (T)), above solid mirrors: a cryogenic beam of H and D with normal velocities down to a few $cm/s$ has been developed \cite{Killian2023,Killian2024} for experiments on QR, whispering-gallery states (WGS), and gravitational quantum states (GQS) of atoms. 
So far, GQS and WGS have been observed with neutrons \cite{Nesvizhevsky2002,Nesvizhevsky2010} and proposed for antihydrogen ($\bar{H}$) \cite{Voronin2011,Voronin2012}; for antiatoms and atoms, their lifetimes and level shifts are set by QR from the mirror.

The solid mirrors for such experiments are not neutral. For example, thermal and native oxides (SiO$_2$) on the silicon (Si) surface carry fixed charges with surface densities of $10^{10}$--$10^{12}~e/cm^2$ \cite{Deal1980,Sze2007}, and ions adsorbed from the residual gas occur on any insulating surface \cite{McGuirk2004,Obrecht2007,Hattermann2012,Herring1949,Speake2003,Behunin2012,Behunin2014}. The field of each charge polarizes the incident atom and adds an attraction to the Casimir--Polder (CP) potential \cite{Casimir1948,Lifshitz1956,DLP1961}. Since QR is formed far from the surface, where the CP potential is weak, an additional attraction even of a modest strength can change the reflection coefficient; for $\bar{H}$ in GQS, this quenching was studied in Ref.~\cite{Voronin2016}. The spatial distribution of the charges is not known in advance: fixed oxide charges and adsorbed ions are located at random positions, charges may be grouped in clusters, or compensated over a finite length by charges of the opposite sign or carriers in the substrate, and charges deposited on purpose may be ordered.

The aim of this paper is to investigate theoretically how charges on the dielectric surface, with these types of spatial distribution, change the QR of H isotopes. We calculate the coefficient of specular reflection and the probability of diffuse scattering of H, D, and T as a function of the density of the charges and of the structure of their distribution. The atoms, velocities, and materials are chosen so that, on the one hand, the charges produce qualitatively new and quantitatively significant effects, and, on the other hand, the conditions are within the reach of the planned experiments.

H isotopes are the natural choice for this problem. For the retarded CP potential $-C_4/z^4$, the reflection is formed on the length $\beta_4=(2mC_4)^{1/2}/\hbar$, and the threshold regime extends up to the normal velocity $v_A=\hbar/m\beta_4\propto m^{-3/2}$: for H above Si, it is 2.8~m/s, while for sodium it is only a few $mm/s$. Light atoms are therefore reflected efficiently at normal velocities accessible with beams. The polarizability of H is known exactly, so the CP potential is fixed by the optical response of the mirror alone. The three isotopes differ only in mass, which changes the strength of the charge effect relative to the CP one and allows the two to be separated.

 The source of H or D at a temperature of $\sim$6~K and the parabolic beam collimation \cite{Killian2023,Killian2024} can provide a broad distribution of longitudinal velocities, with maxima at $\sim$310~m/s for H and $\sim$220~m/s for D, respectively. We consider the range of 100--300~m/s to be accessible for the experiments; the flux at lower velocities is too low and the required incidence angle for higher velocities is too small. At these velocities, the atom passes over a charge much faster than the reflection region responds, so that the effect of the charges on the specular wave does not depend on the longitudinal velocity, and the results apply to the whole velocity spread of the beam (Sec.~\ref{sec:diag}). Normal velocities, determined by grazing angles of $10^{-4}$--$10^{-2}$~rad, are 3--100~cm/s. This range covers the threshold regime, in which the charges change the velocity dependence of the reflection qualitatively and their effect is largest, and its upper end, where the reflection is formed inside the CP region and the charge effect becomes small. At lower normal velocities, the motion above a horizontal mirror is quantized by gravity.

The two materials considered here represent the two classes of mirrors and the two limits of the charge effect. Si covered by its native oxide, SiO$_2$(2~nm)/Si, is a standard material for QR experiments \cite{Shimizu2001,Pasquini2004,Oberst2005}; its large permittivity, $\varepsilon=11.7$, weakens the field of the charges, and the charge effect is moderate. Magnesium fluoride, MgF$_2$, is a wide-gap dielectric with $\varepsilon=5.2$, on which the charges are not screened; the coupling to the charges is 4.2 times stronger and for T it approaches the critical value at which the threshold law is lost. For other dielectric mirrors, the charges enter through the factor $(1+\varepsilon)^{-2}$ of the coupling, Eq.~(\ref{eq:zeta}): relative to Si, it is 7.0 for fused silica, 2.7 for CaF$_2$, and 1.0--1.5 for sapphire. The charge densities of $10^{10}$--$10^{12}$~cm$^{-2}$ cover oxides of different quality, from densities whose effect is below the precision of the experiments to densities that change the absorption by tens of percent or more.

The following theory uses the condition of complete absorption at the surface: an atom that has passed the reflection region and reached the surface leaves the specular channel due to incoherent contact scattering on disorders of the surface on the atomic scale.  For $\bar{H}$, this condition is provided by annihilation \cite{Voronin2005,Dufour2013}, but for ordinary atoms it is not self-evident;  we justify it for solid surfaces in Sec.~\ref{sec:absorbjust}.

Our approach is as follows. The polarization potential of the charges varies along the mirror surface, and its Fourier components couple the specular wave to channels in which the atom is scattered off specularly. At grazing incidence, these channels are detuned from the specular one by two to three orders of magnitude more than the energy resolution of the reflection region, so that the specular wave responds only to the potential averaged over the mirror surface, whereas the lateral variations produce diffuse scattering. For uncorrelated charges, the averaged potential is an inverse-square tail $-C_2/z^2$, and specular reflection becomes the radial problem of a free particle with a non-integer angular momentum, whose exact solution is known \cite{LandauQM,Case1950,Frank1971}. The central result is that the charges change the QR threshold law from linear $1-\Rr\propto\vperp$, to a fractional power $\vperp^{2\nu}$ with an exponent fixed by the charge density, which for the densities of interest appears as a logarithmic velocity dependence of the charge-induced absorption.

The paper is organized as follows. Section~\ref{sec:neutral} recalls the reflection from the neutral mirror, defines the length and time scales of the reflection region, and justifies the condition of complete absorption. Section~\ref{sec:charges} introduces the charges, writes the exact channel equations, and derives the closed equation for the specular wave together with the averaged potential and its structure factor. Section~\ref{sec:specular} solves the specular problem, gives the threshold law, its logarithmic expansion, and the absorption coefficients of the three isotopes at both types of mirrors, and discusses the isotope effect, the effect of charge correlations, and the strong-coupling regime. Section~\ref{sec:diffuse} estimates diffuse scattering, and Sec.~\ref{sec:concl} summarizes the results and their consequences for experiments. Formulas are written in Gaussian units; numerical values are given in atomic units, unless units are stated.

\begin{figure}[t]
\includegraphics[width=\columnwidth]{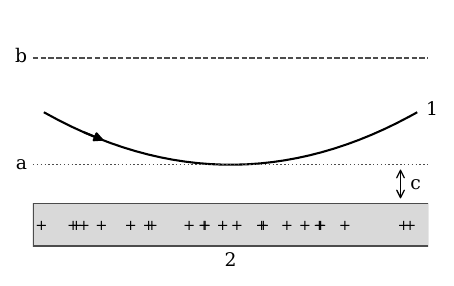}
\caption{Geometry of the problem. An atom (trajectory 1) moves along the mirror (2) with the longitudinal velocity $\vpar$ much larger than the normal velocity $\vperp$. The reflection is formed at heights of the order of the absorption length $\beta_4$ (dotted line a, arrow c), Eq.~(\ref{eq:beta4}); the inverse-square tail of the averaged charge potential acts up to the height $z_{\max}$ of Eq.~(\ref{eq:zmax}) (dashed line b). Crosses: surface charges.}
\label{fig:geometry}
\end{figure}

\section{QR from the flat neutral mirror}
\label{sec:neutral}

\subsection{Equation and boundary condition}

We consider the ideal flat mirror which surface absorbs atoms completely as a starting point of our formalism. We will show  further that this model is valid for a wide class of solid surfaces. The atom-surface potential in such a simplified case depends only on the height $z$ above the mirror, the motion along the mirror is free, and the wave function is the product of a plane wave along the mirror and a function $u(z)$, which for an atom of mass $m$ obeys:
\begin{equation}
-\frac{\hbar^2}{2m}\,u'' + V_{\rm CP}(z)\,u = \Eperp u ,
\label{eq:schr0}
\end{equation}
where $\Eperp=\hbar^2k^2/2m$ is the part of the kinetic energy associated with the motion towards the mirror; it is conserved separately because the potential does not depend on the position along the mirror. We consider the two materials introduced in Sec.~\ref{sec:intro}, SiO$_2$(2~nm)/Si and MgF$_2$. For $V_{\rm CP}$, we use the Lifshitz potentials described in Appendix~\ref{app:lifshitz}. The asymptotes of $V_{\rm CP}$ are
\begin{align}
V_{\rm CP}(z) &\simeq -\frac{C_3}{z^3}\quad (z\ll z^*), \nonumber\\
V_{\rm CP}(z) &\simeq -\frac{C_4}{z^4}\quad (z\gg z^*), \qquad z^*=\frac{C_4}{C_3},
\label{eq:asympt}
\end{align}
with $C_3=0.1007$, $C_4=50.28$, and $z^*=26$~nm for SiO$_2$/Si and $C_3=0.0466$, $C_4=38.76$, and $z^*=44$~nm for MgF$_2$. The masses of H, D, and T are 1837.15, 3670.48, and 5496.92 electron masses.

An atom that is not reflected in the CP region reaches the surface and is lost. We will show in Sect.~\ref{sec:absorbjust} that for most realistic surfaces, the atom leaves the specular channel due to incoherent scattering on atomic scale disorders. This phenomenon is described by the boundary condition of complete absorption: only the wave incoming towards the surface is present near it,
\begin{equation}
u(z)\to \frac{\exp\big[-i\!\int^{z}\! k_{\rm loc}(z',\Eperp)\,\dd z'\big]}{\sqrt{k_{\rm loc}(z,\Eperp)}}
\quad (z\to0),
\label{eq:absorb}
\end{equation}
with the local wave number
\begin{equation}
k_{\rm loc}(z,\Eperp)=\frac{\sqrt{2m[\Eperp-V_{\rm CP}(z)]}}{\hbar}.
\label{eq:kloc}
\end{equation}
The condition is well defined because the semiclassical approximation becomes exact in the van der Waals region as the height tends to zero, so that the reflection amplitude does not depend on where the condition is imposed \cite{Voronin2003}; Appendix~\ref{app:numerics} shows the numerical convergence.

\subsection{Threshold law and scales of the reflection region}

At zero energy, the solution of Eqs.~(\ref{eq:schr0}) and (\ref{eq:absorb}) behaves at large heights as
\begin{equation}
u(z)\simeq z-a, \qquad a=a'-i a'',
\label{eq:scatlength}
\end{equation}
which defines the complex scattering length. At the normal wave number
\begin{equation}
k=\frac{m\vperp}{\hbar},
\label{eq:k}
\end{equation}
the solution of Eqs.~(\ref{eq:schr0}) and (\ref{eq:absorb}) has, outside the CP region, the asymptotic form
\begin{equation}
\chi(z)=e^{-ika(k)}\,\frac{\sin k[z-a(k)]}{k}\qquad (z\gg\beta_4),
\label{eq:chi}
\end{equation}
where the ratio of the incoming and outgoing amplitudes defines the energy-dependent effective scattering length
\begin{equation}
a(k)=a+O(k),\qquad a(0)=a,
\label{eq:ak}
\end{equation}
and the exact reflection coefficient is
\begin{equation}
\Rr=e^{4k\operatorname{Im}a(k)} .
\label{eq:Rak}
\end{equation}
The threshold law keeps only the energy-independent term of Eq.~(\ref{eq:ak}),
\begin{equation}
\Rr\simeq e^{-4ka''}\simeq 1-4ka'' ,
\label{eq:threshold}
\end{equation}
and is valid as long as the correction to $a(k)$ is small, i.e.\ when the wavelength exceeds the size of the reflection region (Eq.~(\ref{eq:vA}) below).

The characteristic length of the reflection follows from the pure retarded potential $-C_4/z^4$, for which the zero-energy equation has the exact solution
\begin{equation}
u(z)=z\exp\Big(\frac{i\beta_4}{z}\Big), \qquad \beta_4=\frac{\sqrt{2mC_4}}{\hbar},
\label{eq:beta4}
\end{equation}
with the scattering length
\begin{equation}
a=-i\beta_4 .
\label{eq:apure}
\end{equation}
The pure retarded potential thus absorbs over the length $\beta_4$, which is also the height at which the reduced local wavelength equals the height. Table~\ref{tab:param} gives $\beta_4$ and the scattering lengths of the Lifshitz potentials: for the SiO$_2$/Si mirror, $a''$ exceeds $\beta_4$ by 1.7--4.8\%, and the real part is 22\%, 14\%, and 10\% of the imaginary one for H, D, and T. The MgF$_2$ mirror absorbs less; its imaginary scattering lengths are 40\% smaller.

\begin{figure*}[t]
\includegraphics[width=\textwidth]{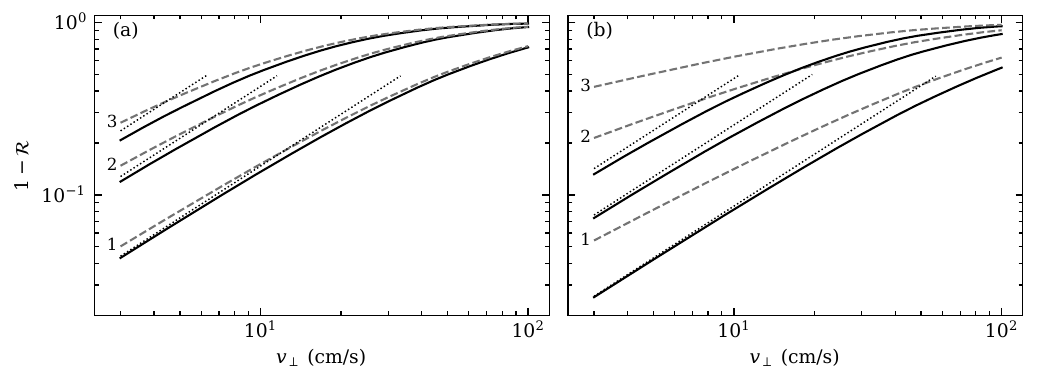}
\caption{Absorption $1-\Rr$ of H (curves 1), D (2), and T (3) as a function of the normal velocity at (a) the SiO$_2$(2~nm)/Si mirror and (b) the MgF$_2$ mirror. Solid black: neutral mirror, exact solution of Eq.~(\ref{eq:schr0}); dashed gray: mirror with $10^{12}$ uncorrelated single charges per cm$^2$, exact solution of Eq.~(\ref{eq:tail}); dotted: threshold law (\ref{eq:threshold}) of the neutral mirror, drawn where it is below 0.5. The curves interpolate the values of Table~\ref{tab:coeff}. The charges shift the curves upward at low velocity, where the threshold law holds, and have little effect at high velocity, where the reflection is formed inside the CP region.}
\label{fig:absorption}
\end{figure*}

The threshold law (\ref{eq:threshold}) thus requires
\begin{equation}
k\beta_4\ll1 \quad\Longleftrightarrow\quad \vperp\ll v_A\equiv\frac{\hbar}{m\beta_4},
\label{eq:vA}
\end{equation}
where $v_A$ is listed in Table~\ref{tab:param}; above this velocity, the energy dependence of $a(k)$ becomes essential and the exact solution is used. Figure~\ref{fig:absorption} compares the threshold law with the exact solution of Eq.~(\ref{eq:schr0}) for both materials (solid and dotted curves), and the columns $N=0$ of Table~\ref{tab:coeff} give the exact values.

The response of the normal motion in the reflection region is characterized by the phase time, i.e.\ the energy derivative of the phase accumulated between the surface and $\beta_4$ on the way in and out,
\begin{align}
\tau_r&=2\hbar\frac{\dd}{\dd\Eperp}\int_0^{\beta_4}\! k_{\rm loc}(z,\Eperp)\,\dd z\Big|_{\Eperp=0}
\nonumber\\
&=2m\!\int_0^{\beta_4}\!\!\frac{\dd z}{\hbar k_{\rm loc}(z,0)} ,
\label{eq:taur}
\end{align}
which, for the pure retarded potential, equals
\begin{equation}
\tau_r=\frac{2m\beta_4^2}{3\hbar}.
\label{eq:taur4}
\end{equation}
Its inverse sets the energy resolution of the reflection region: perturbations whose detuning exceeds $\hbar/\tau_r$ act only through virtual excursions. The values for the Lifshitz potential are listed in Table~\ref{tab:param}.

\subsection{Assumption of complete absorption}
\label{sec:absorbjust}

\paragraph*{Partial reflection by the wall.}
Complete absorption means that the wave reflected by the repulsive wall does not return to the specular channel coherently. Let the wall reflect the incoming wave with the amplitude $\rho$, $|\rho|\le1$, defined at a height $z_0$ a few \aa ngstr\"om above the wall, where the semiclassical approximation holds:
\begin{equation}
u(z)\simeq\frac{e^{-i\Phi(z)}+\rho\,e^{\,i\Phi(z)}}{\sqrt{k_{\rm loc}(z,\Eperp)}},\qquad
\Phi(z)=\int_{z_0}^{z}k_{\rm loc}\,\dd z',
\label{eq:absorbrho}
\end{equation}
for $z\simeq z_0$; $\rho=0$ is the condition (\ref{eq:absorb}). Wave-function $u$ is a combination of semiclasicall incoming and partially reflected waves, $\Phi_\infty$ is the phase accumulated by the wave between $z_0$ and the asymptotic region.
At $z\gg\beta_4$
\begin{equation}
u \to C\,(z-a),
\label{eq:Phiinf}
\end{equation}
where $a$ is the scattering length.  Matching inner and asymptotic  solutions gives expression for a scattering length in terms of "full absorption" scattering length $a_0$, contact partial reflection amplitude $\rho$ and accumulated phase $\Phi$:
\begin{equation}
a(\tilde\rho)=\frac{a_0+\tilde\rho\,a_0^*}{1+\tilde\rho},\qquad
\tilde\rho=\rho\,e^{2i\Phi_\infty},
\label{eq:arho}
\end{equation}

For pure retarded potential and total absorption, the boundary condition $u =z\,e^{-i\Phi(z)}/\sqrt{\beta_4}$ is exact, with $a_0=-i\beta_4$ and $\Phi_\infty=\beta_4/z_0$ ($\approx70$ for H at $z_0=6\,a_0$).
\begin{equation}
\begin{gathered}
a''(\tilde\rho)=a_0''\,\frac{1-|\tilde\rho|^2}{|1+\tilde\rho|^2},\\
-\frac{2|\tilde\rho|}{1+|\tilde\rho|}\le\frac{a''(\tilde\rho)-a_0''}{a_0''}\le\frac{2|\tilde\rho|}{1-|\tilde\rho|} .
\end{gathered}
\label{eq:a2rho}
\end{equation}
The amplitude $\tilde\rho$ is the wall reflection referred to the asymptotic region. It does not depend on the choice of $z_0$: a shift of $z_0$ by a phase $\delta$ of $\Phi$ multiplies $\rho$ by $e^{2i\delta}$ and $e^{2i\Phi_\infty}$ by $e^{-2i\delta}$. Since $k_{\rm loc}\propto m^{1/2}$, $\Phi_\infty$ scales as $m^{1/2}$. Equations (\ref{eq:arho}) and (\ref{eq:a2rho}) hold for any real potential; for the Lifshitz potential, they agree with the numerical solution. 

\paragraph*{Averaging of the contact amplitude.}
On a real surface, the contact reflection depends on the lateral position $\bm{R}$. It is linear in the wave function, $\psi_{\rm out}(\bm{R})=\int\rho(\bm{R},\bm{R}')\,\psi_{\rm in}(\bm{R}')\,\dd^2R'$, so the amplitudes reflected by different sites add. For the incident lateral wave $e^{i\bm{K}\cdot\bm{R}}$, the component reflected with the same lateral wave vector has the amplitude
\begin{equation}
\langle\rho\rangle=\frac{1}{S}\iint_S e^{-i\bm{K}\cdot\bm{R}}\,\rho(\bm{R},\bm{R}')\,e^{i\bm{K}\cdot\bm{R}'}\,\dd^2R\,\dd^2R' ,
\label{eq:rhoavg}
\end{equation}
which reduces to $S^{-1}\int_S\rho(\bm{R})\,\dd^2R$ for a local kernel $\rho(\bm{R})\,\delta(\bm{R}-\bm{R}')$. Above $z_0$, the potential is laterally uniform, the lateral Fourier components propagate independently, and the specular component obeys Eqs.~(\ref{eq:schr0}) and (\ref{eq:absorbrho}) with $\rho\to\langle\rho\rangle$. The phase $\Phi_\infty$ is common to all sites, so that
\begin{equation}
a_{\rm spec}=a\big(\langle\tilde\rho\rangle\big),\qquad
\langle\tilde\rho\rangle=e^{2i\Phi_\infty}\langle\rho\rangle,\qquad
|\langle\tilde\rho\rangle|=|\langle\rho\rangle| .
\label{eq:aspec}
\end{equation}
The specular channel is thus fixed by the averaged contact amplitude.

Equation (\ref{eq:aspec}) is based on three conditions.

(i) \emph{Uniformity above $z_0$.} A lateral Fourier component $G=2\pi/a_s$ of the potential decays as $e^{-Gz}$. For $a_s=2.5$--5~\AA\ and $z_0=6\,a_0$, the corrugation is $e^{-Gz_0}=3\times10^{-4}$ to $2\times10^{-2}$ of the potential, and it vanishes at the reflection height $z\sim\beta_4$.

(ii) \emph{Loss of the remainder.} The remainder $\rho-\langle\rho\rangle$ carries lateral wave numbers $q\sim2\pi/a_s$, and
\begin{equation}
\frac{\hbar^2q^2}{2m}=3\text{--}13~\text{meV}\ \gg\ \frac{m\vpar^2}{2}=0.05\text{--}0.5~\text{meV}
\label{eq:closed}
\end{equation}
for $\vpar=100$--300~m/s. These channels are closed and decay within $1/q\le0.8$~\AA\ of the wall; the atom in them is bound in the physisorption well and loses coherence by phonon emission. Open channels with small $q$ leave the surface in other directions and, above $z_0$, do not couple to the specular one.

(iii) \emph{Averaging area.} The reflection region resolves the lateral energy to $\hbar/\tau_r$, Eq.~(\ref{eq:taur}). A transfer $q_x$ along the motion changes it by $\hbar\vpar q_x$, a transfer $q_y$ across the motion by $\hbar^2q_y^2/2m$. The specular channel therefore collects the amplitudes from the area
\begin{equation}
S=L_xL_y,\qquad L_x=\vpar\tau_r,\qquad L_y=\Big(\frac{\hbar\tau_r}{2m}\Big)^{1/2},
\label{eq:Sarea}
\end{equation}
provided the atomic wave is coherent over it. Table~\ref{tab:param} gives $L_x=0.8$--19~$\mu$m and $L_y=16$--26~nm (H to T), so that $S\ge1.3\times10^{4}$~nm$^2$, far smaller than the coherence area of a collimated beam (Sec.~\ref{sec:averaged}). The area contains $N=S/a_s^2\ge5\times10^4$ sites, and the average (\ref{eq:rhoavg}) differs from the mean over the surface by
\begin{equation}
|\delta\rho|\lesssim N^{-1/2}\le5\times10^{-3}.
\label{eq:drho}
\end{equation}
This difference changes randomly from atom to atom; its linear contribution to $a''$ cancels in the beam, and the quadratic one is below $2|\delta\rho|^2=5\times10^{-5}$.

\paragraph*{Model of the wall.}
We will show that $|\frac{a''(\tilde\rho)-a_0''}{a_0''}|\ll1$ for a wide class of solid surfaces. A physisorption well of depth $D$ ends in a hard wall at the height $h(\bm{R})+u_z$, where $h$ is the static height relative to the laterally uniform potential, $\langle h\rangle=0$, and $u_z$ is the vibrational displacement. The CP potential at 3~\AA\ gives $D=15$~meV for SiO$_2$/Si and 6.9~meV for MgF$_2$; H is bound to graphite by 39~meV \cite{Ghio1980}. We take $D=15$--40~meV for SiO$_2$/Si and, scaled by the van der Waals constants, 6.9--18.5~meV for MgF$_2$. At the bottom of the well
\begin{equation}
k_w=\frac{\sqrt{2mD}}{\hbar}=1.8\text{--}4.4~\text{\AA}^{-1}.
\label{eq:kw}
\end{equation}
The contact, two crossings of a well of width $\ell_w\approx3$~\AA, lasts
\begin{equation}
\tau_w=\frac{2\ell_w m}{\hbar k_w}=0.2\text{--}0.5~\text{ps},\qquad \vpar\tau_w=0.2\text{--}1.6~\text{\AA}<a_s ,
\label{eq:tauw}
\end{equation}
so that the kernel is local. In the eikonal approximation, a site reflects as a flat wall displaced by $h+u_z$, which lengthens the path in the well by $2(h+u_z)$:
\begin{equation}
\rho(\bm{R})=\rho_w\,e^{2ik_w[h(\bm{R})+u_z]},\qquad|\rho_w|\le1,
\label{eq:rhoR}
\end{equation}
and, since $h$ and $u_z$ are independent,
\begin{equation}
\langle\rho\rangle=\rho_w\,\big\langle e^{2ik_wu_z}\big\rangle\,\big\langle e^{2ik_wh}\big\rangle .
\label{eq:rhofact}
\end{equation}
The vibrational factor, which we account here for completeness, is the Debye--Waller factor $e^{-W}$, $W=2k_w^2\langle u_z^2\rangle$ : 
0.84--0.97 at 4~K and 0.64--0.94 at 300~K for atoms of 16--19~u with $\theta_D\approx500$~K \cite{Farias1998}. The phonon-exchange part of the wave is lost from the specular channel. 

For the justification of boundary condition of total absorption, the static factor is decisive: a displacement of $\pi/k_w=0.7$--1.7~\AA\ changes the phase by $2\pi$. For a disordered surface, we take $h$ Gaussian with the rms value $\sigma_s$, and $\langle e^{iX}\rangle=e^{-\langle X^2\rangle/2}$ gives
\begin{equation}
\big\langle e^{2ik_wh}\big\rangle=e^{-2k_w^2\sigma_s^2}.
\label{eq:static}
\end{equation}
The outermost atoms are spread in height over the interatomic distances with exception of atomic ordered surfaces. In particular, $b=1.6$--2.6~\AA\ (Si--O to O--O) for amorphous oxide and 2.0--2.6~\AA\ (Mg--F to F--F) for the  surface layer of polished MgF$_2$ \cite{Randi2005}. Hence,
\begin{equation}
\sigma_s\simeq\frac{b}{\sqrt{12}}=0.46\text{--}0.75~\text{\AA},\qquad
2k_w^2\sigma_s^2=2.2\text{--}22,
\label{eq:sigmareal}
\end{equation}
and, even for $|\rho_w|=1$,
\begin{equation}
|\langle\rho\rangle|\le e^{-W}e^{-2k_w^2\sigma_s^2}=0.1\ \text{to}\ 3\times10^{-10}.
\label{eq:rhonum}
\end{equation}
Table~\ref{tab:rho} gives the resulting bounds (\ref{eq:a2rho}). The exponent grows in proportion to the mass, and the deviation from complete absorption is below 9\% (H), 0.4\% (D), and 0.02\% (T) on SiO$_2$/Si and below 24\%, 2.3\%, and 0.25\% on MgF$_2$; the H bounds are reached for the shallow well and the smallest disorder.

\begin{table}[!htb]
\caption{Upper bound of the deviation $\delta a''/a''$ from complete absorption, Eqs.~(\ref{eq:a2rho}) and (\ref{eq:rhonum}), at 4~K and for $|\rho_w|=1$. (a) H, for the well depth $D$ and the rms static height disorder $\sigma_s$. (b) The mirrors of this paper: SiO$_2$/Si with $D=15$--40~meV and $\sigma_s=0.46$--0.75~\AA, MgF$_2$ with $D=6.9$--18.5~meV and $\sigma_s=0.58$--0.75~\AA.}
\label{tab:rho}
\begin{ruledtabular}
\begin{tabular}{lccc}
\multicolumn{4}{l}{(a) Hydrogen}\\
$\sigma_s$ (\AA) & & $D=15$~meV & $D=40$~meV \\
\hline
0.3  & & $<70\%$ & $<6\%$ \\
0.46 & & $<10\%$ & $<0.05\%$ \\
0.75 & & $<0.06\%$ & $<10^{-7}\%$ \\
1.0  & & $<10^{-4}\%$ & $<10^{-7}\%$ \\
\hline
\multicolumn{4}{l}{(b) Mirrors of this paper}\\
 & H & D & T \\
\hline
SiO$_2$/Si & $<9\%$ & $<0.4\%$ & $<0.02\%$ \\
MgF$_2$ & $<24\%$ & $<2.3\%$ & $<0.25\%$ \\
\end{tabular}
\end{ruledtabular}
\end{table}

\paragraph*{ Roughness and ordered surfaces.}

On an ordered surface with the corrugation $h=h_0\cos(\bm{G}\cdot\bm{R})$
\begin{equation}
\big\langle e^{2ik_wh}\big\rangle=J_0(2k_wh_0),
\label{eq:J0}
\end{equation}
which is 0.4--0.9 for $h_0=0.1$--0.2~\AA\ \cite{Hoinkes1980}, and complete absorption fails. Roughness on lateral scales larger than the reflection height shifts the wall and the long-range potential together. It leaves $\tilde\rho$ unchanged and multiplies the reflection amplitude by $e^{2ikh}$, with $k$ of Eq.~(\ref{eq:k}):
\begin{equation}
2kh\le3\times10^{-2}\qquad(h=1~\text{nm},\ \vperp\le1~\text{m/s}).
\label{eq:polish}
\end{equation}
Complete absorption thus holds for amorphous oxides and polished crystals and might fail for atomically ordered surfaces and for superfluid helium, which binds hydrogen by only 1~K \cite{Doyle1991,Yu1993,Crepin2017,Crepin2019,Karakhanyan2026}.

\paragraph*{Isotope test.}
The phases $2\Phi_\infty$ and $2k_wh$ scale as $m^{1/2}$ and are many radians, so the phases of $\langle\tilde\rho\rangle$ for H, D, and T are unrelated. By Eq.~(\ref{eq:a2rho}) the ratio $r_{\rm CP}=m_{\rm X}a''_{\rm X}/m_{\rm H}a''_{\rm H}$ of the threshold absorptions of the isotope X and of H lies within
\begin{equation}
\Big(\frac{1-|\langle\rho\rangle|}{1+|\langle\rho\rangle|}\Big)^2\le\frac{r_{\rm CP}}{r_{\rm CP}^{(0)}}\le\Big(\frac{1+|\langle\rho\rangle|}{1-|\langle\rho\rangle|}\Big)^2,
\label{eq:isotest}
\end{equation}
where $r_{\rm CP}^{(0)}$ is its value for complete absorption.

\section{Surface charges and the specular equation}
\label{sec:charges}

\subsection{Charges and their field}

We assume that the charges are not compensated on heights relevant for reflection, from 10~nm to micrometers, or are compensated on a comparable or larger length only. This holds when no mobile carriers screen them: for wide-gap dielectric materials such as MgF$_2$, for Si at cryogenic temperature, where the carriers are frozen out, and, with partial compensation within about 0.3~$\mu$m, for high-purity Si at room temperature. On a conducting substrate, each charge forms a dipole with its image, and the effects described below are absent.

The charges are fixed at the positions $\bX_j$ in the mirror plane, with the charge numbers $Q_j$. Their surface density and its Fourier components are
\begin{equation}
\sigma(\br)=\sum_j Q_j\,\delta^{(2)}(\br-\bX_j),\qquad
\sigma_{\bq}=\sum_j Q_j\,e^{-i\bq\cdot\bX_j},
\label{eq:sigma}
\end{equation}
where $\br$ is the in-plane position, and the Fourier convention for any function of $\br$ on the mirror area $S$ is
\begin{equation}
f(\br)=\frac1S\sum_{\bq}f_{\bq}\,e^{i\bq\cdot\br},\qquad
\sum_{\bq}\longrightarrow S\!\int\!\frac{\dd^2q}{(2\pi)^2}.
\label{eq:fourier}
\end{equation}
We denote by $N$ the number of elementary charges per unit area, and by $n_c$ the number of clusters (groups of $Q_c$ charges) per unit area, so that $N=n_cQ_c$.

A point charge $Qe$ at the boundary between vacuum and a medium of static permittivity $\varepsilon$ acts in vacuum as charge \cite{Jackson}
\begin{equation}
Q_{\rm eff}=\frac{2Qe}{1+\varepsilon},
\label{eq:Qeff}
\end{equation}
with the static permittivity $\varepsilon=11.7$ for Si, which determines the field at heights much larger than the oxide thickness, and $\varepsilon=5.2$ for MgF$_2$. With the identity
\begin{equation}
\int\!\dd^2 r\,\frac{e^{-i\bq\cdot\br}}{\sqrt{r^2+z^2}}=\frac{2\pi}{q}\,e^{-qz},
\label{eq:identity}
\end{equation}
the Fourier components of the field of the charges are
\begin{equation}
\tilde{\bE}_{\bq}(z)=\frac{4\pi e}{1+\varepsilon}\,\sigma_{\bq}\,e^{-qz}\,(\hz-i\hq),
\label{eq:field}
\end{equation}
with $\hq$ the unit vector along $\bq$. The field at height $z$ is thus formed by the charge distribution on lateral scales larger than $z$.

The field induces the dipole moment $\alpha_p\bE$, with the static polarizability $\alpha_p=4.5\,a_0^3$ for all isotopes, and the interaction \cite{LandauECM}
\begin{equation}
V_{\rm ch}(\br,z)=-\frac{\alpha_p}{2}\,|\bE(\br,z)|^2 .
\label{eq:Vch}
\end{equation}
With the convention (\ref{eq:fourier}), its Fourier components are
\begin{equation}
\tilde V_{\bq}(z)=\frac1S\!\int\!\dd^2r\,e^{-i\bq\cdot\br}\,V_{\rm ch}(\br,z)
=-\frac{\alpha_p}{2S^2}\sum_{\bq'}\tilde{\bE}_{\bq'+\bq}\cdot\tilde{\bE}^*_{\bq'} .
\label{eq:Vq}
\end{equation}

\subsection{Channel equations}

The atom moves with the longitudinal wave vector
\begin{equation}
\bK=\frac{m\bm{v}_\parallel}{\hbar}
\label{eq:K}
\end{equation}
in the CP potential and the potential of the charges. Expanding the wave function in the plane waves generated by the charge potential,
\begin{equation}
\Psi(\br,z)=\sum_{\bq}e^{i(\bK+\bq)\cdot\br}\,\phi_{\bq}(z),
\label{eq:expansion}
\end{equation}
gives the exact channel equations
\begin{equation}
\Big[-\frac{\hbar^2}{2m}\frac{\dd^2}{\dd z^2}+V_{\rm CP}+\tilde V_0-\epsilon_{\bq}\Big]\phi_{\bq}
+\sum_{\bq'\neq\bq}\tilde V_{\bq-\bq'}\,\phi_{\bq'}=0,
\label{eq:channels}
\end{equation}
with the channel energies
\begin{equation}
\epsilon_{\bq}=\frac{\hbar^2k^2}{2m}-\frac{\hbar^2(2\bK\cdot\bq+q^2)}{2m}.
\label{eq:epsq}
\end{equation}
The channel with zero wave vector is the specular one; the diagonal element $\tilde V_0$ is the charge potential averaged over the mirror plane. Equation (\ref{eq:channels}) is exact for one frozen configuration $\{\bX_j\}$ and the absorbing condition (\ref{eq:absorb}) is imposed in every channel.

\subsection{Closed equation for the specular wave}
\label{sec:diag}

Eliminating the channels $\bq\neq0$ from Eq.~(\ref{eq:channels}) to second order in the off-diagonal elements gives the specular equation
\begin{equation}
\Big[-\frac{\hbar^2}{2m}\frac{\dd^2}{\dd z^2}+V_{\rm CP}+\tilde V_0-\frac{\hbar^2k^2}{2m}\Big]\phi_0
-\sum_{\bq\neq0}\tilde V_{-\bq}\,G_{\bq}\,\tilde V_{\bq}\,\phi_0=0,
\label{eq:second}
\end{equation}
where $G_{\bq}$ is the Green function of channel $\bq$. The channels are detuned from the specular one by
\begin{equation}
\epsilon_0-\epsilon_{\bq}=\hbar\vpar q_x,
\label{eq:detuning}
\end{equation}
with the $x$ axis along the beam, where the quadratic term of Eq.~(\ref{eq:epsq}) has been dropped; its ratio to the linear one is
\begin{equation}
\frac{q}{2K}=\frac{\hbar}{2m\vpar\beta_4}\qquad (q=1/\beta_4),
\label{eq:quadratic}
\end{equation}
which is below $1.4\times10^{-2}$ for the beams considered.

Three conditions make the last term in Eq.~(\ref{eq:second}) negligible. (i) The Fourier components that act at the reflection height have wave numbers up to $1/\beta_4$ [Eq.~(\ref{eq:field})], and their detuning must exceed the energy resolution $\hbar/\tau_r$ of the reflection region. This is expressed by the adiabaticity parameter
\begin{equation}
\theta=\frac{\hbar/\tau_r}{\hbar\vpar/\beta_4}=\frac{\beta_4}{\vpar\tau_r},
\label{eq:theta}
\end{equation}
which is below $3\times10^{-2}$ (Table~\ref{tab:param}) and, since it scales as $1/\vpar$, remains below 0.1 down to 30~m/s for H and to lower velocities for D and T. Equivalently, the detuning exceeds the CP potential at the reflection height, $\hbar\vpar/\beta_4\gg|V_{\rm CP}(\beta_4)|$, by a factor of 130--1550, and the diffuse channels lie far above their thresholds: their normal wave numbers,
\begin{equation}
k_q\beta_4=\Big(\frac{2m\vpar\beta_4}{\hbar}\Big)^{1/2}\qquad (q=1/\beta_4),
\label{eq:kq4}
\end{equation}
are of the order of ten (Table~\ref{tab:param}). (ii) When the channels are far off shell they act only through short virtual excursions, and the last term of Eq.~(\ref{eq:second}) reduces to a local potential of the order of the squared potential of one cluster divided by the detuning. Its ratio to the first-order effect of the same cluster is of the order of one half of the phase
\begin{equation}
\varphi_1=\frac{1}{\hbar\vpar}\int_{-\infty}^{\infty}\!|v(x,\beta_4)|\,\dd x
=\frac{\pi\alpha_p}{4\hbar\vpar\beta_4^3}\Big(\frac{2Q_ce}{1+\varepsilon}\Big)^2
\label{eq:phi1}
\end{equation}
that the potential $v$ of one cluster, Eq.~(\ref{eq:vcluster}) below, imprints on the wave during the passage at the reflection height. For single charges, $\varphi_1$ is below $3\times10^{-5}$ on Si and below $1.2\times10^{-4}$ on MgF$_2$; it grows as $Q_c^2$ and reaches $10^{-1}$ only for clusters of about 60 charges on Si at 100~m/s. (iii) The averaged potential $\tilde V_0$ must be formed by many clusters at the reflection height. The potential at the height $z$ is formed by the clusters within a lateral distance of the order of $z$ (Eq.~(\ref{eq:field})), and during the response time $\tau_r$ of the reflection region the atom at the height $\beta_4$ travels the distance $\vpar\tau_r$ along the mirror, which is 1.7, 6, and 13~$\mu$m for H, D, and T at 200~m/s. The number of clusters it sees during this time,
\begin{equation}
N_\times=2\beta_4\,n_c\,\vpar\,\tau_r ,
\label{eq:Ncross}
\end{equation}
is about 750, 3900, and 9900 for the three isotopes at $10^{12}$~$e/cm^2$ and scales with the density; even at $10^{10}$~e/cm$^{2}$ it exceeds 7 (H) and 40 (D). The specular wave therefore responds to the accumulated potential of many clusters, with a relative fluctuation of the order of $N_\times^{-1/2}$.

Under these conditions, we drop the off-diagonal elements in the specular channel and obtain the closed equation
\begin{equation}
\Big[-\frac{\hbar^2}{2m}\frac{\dd^2}{\dd z^2}+V_{\rm CP}(z)+\tilde V_0(z)-\frac{\hbar^2k^2}{2m}\Big]\phi_0=0 .
\label{eq:specular}
\end{equation}
It does not depend on the longitudinal velocity: the dependence on $\vpar$ resides entirely in the off-diagonal elements, which produce diffuse scattering (Sec.~\ref{sec:diffuse}). In the opposite limit $\theta\gg1$, not realized here, the channels would degenerate with the specular one and the atom would follow the local potential of each cluster adiabatically.
\begin{table}[t]
\caption{Parameters of the three isotopes for the 
Si$_2$(2~nm) /Si mirror and, in the last rows, for the MgF$_2$ mirror: scattering length $a$ of the Lifshitz potential, absorption length $\beta_4$ (Eq.~(\ref{eq:beta4})), velocity $v_A$ (Eq.~(\ref{eq:vA})), phase time $\tau_r$ (Eq.~(\ref{eq:taur})), adiabaticity parameter $\theta$ (Eq.~(\ref{eq:theta})) and normal wave number of the diffuse channels (Eq.~(\ref{eq:kq4})) at longitudinal velocities of 100 and 300~m/s, coupling $\zeta$ (Eq.~(\ref{eq:zeta})) for $10^{12}$$e$/cm$^2$, and threshold constant $\mathcal{C}$ (Eq.~(\ref{eq:log})).
}
\label{tab:param}
\begin{ruledtabular}
\begin{tabular}{lddd}
 & \multicolumn{1}{c}{H} & \multicolumn{1}{c}{D} & \multicolumn{1}{c}{T} \\
\hline
$a'$ (nm)                       & -5.06 & -4.64 & -4.22 \\
$a''$ (nm)                      & 23.13 & 33.54 & 41.23 \\
$\beta_4$ (nm)                  & 22.74 & 32.15 & 39.34 \\
$v_A$ (m/s)                     & 2.77 & 0.98 & 0.54 \\
$\tau_r$ (ns)                   & 8.3 & 30 & 63 \\
$\theta$ (100 m/s)              & 0.028 & 0.011 & 0.0063 \\
$\theta$ (300 m/s)              & 0.0092 & 0.0036 & 0.0021 \\
$k_q\beta_4$ (100 m/s)          & 8.5 & 14.3 & 19.3 \\
$k_q\beta_4$ (300 m/s)          & 14.7 & 24.7 & 33.5 \\
$\zeta$ ($10^{12}$ cm$^{-2}$)   & 0.018 & 0.036 & 0.054 \\
$\mathcal{C}$                   & 0.83 & 0.71 & 0.66 \\
\hline
MgF$_2$: $a'$ (nm)              & -4.24 & -4.41 & -4.48 \\
MgF$_2$: $a''$ (nm)             & 13.55 & 20.06 & 24.90 \\
MgF$_2$: $\beta_4$ (nm)         & 19.97 & 28.23 & 34.54 \\
MgF$_2$: $v_A$ (m/s)            & 3.15 & 1.12 & 0.61 \\
MgF$_2$: $\zeta$ ($10^{12}$ cm$^{-2}$) & 0.076 & 0.151 & 0.226 \\
\end{tabular}
\end{ruledtabular}
\end{table}
\subsection{Averaged potential and structure factor}
\label{sec:averaged}

Setting the transfer to zero in Eq.~(\ref{eq:Vq}) and using Eqs.~(\ref{eq:fourier}) and (\ref{eq:field}) gives
\begin{equation}
\tilde V_0(z)=-\alpha_p\Big(\frac{4\pi e}{1+\varepsilon}\Big)^2\!\int\!\frac{\dd^2q}{(2\pi)^2}\,
\frac{|\sigma_{\bq}|^2}{S}\,e^{-2qz},
\label{eq:V0}
\end{equation}
where $S$ is now the area over which the wave of one atom is coherent, and the uniform field of a net charge, which adds a constant, is omitted (its effect is discussed in Sec.~\ref{sec:diffuse}). The squared modulus separates into single clusters and pairs,
\begin{equation}
|\sigma_{\bq}|^2=\sum_j Q_j^2+\sum_{j\neq l}Q_jQ_l\,e^{-i\bq\cdot(\bX_j-\bX_l)} .
\label{eq:sigma2}
\end{equation}
The single-cluster part does not depend on the positions and gives exactly
\begin{equation}
\tilde V_0^{\rm self}(z)=-\frac{C_2}{z^2},\qquad
C_2=\frac{2\pi\alpha_p e^2\,n_cQ_c^2}{(1+\varepsilon)^2}.
\label{eq:C2}
\end{equation}
It is the sum over the clusters of the lateral average of the potential of one cluster,
\begin{equation}
v(\rho,z)=-\frac{\alpha_p}{2}\Big(\frac{2Q_ce}{1+\varepsilon}\Big)^2\frac{1}{(\rho^2+z^2)^2},\qquad
\rho=|\br-\bX_j|,
\label{eq:vcluster}
\end{equation}
or, for an atom moving along a straight line, the sum over the impact parameters $b$ of the clusters with the uniform weight $n_c$:
\begin{equation}
n_c\!\int_{-\infty}^{\infty}\!\dd b\!\int_{-\infty}^{\infty}\!\dd x\;v\big(\sqrt{x^2+b^2},z\big)=-\frac{C_2}{z^2}.
\label{eq:impact}
\end{equation}
The pair part contains the configuration,
\begin{equation}
\tilde V_0^{\rm pair}(z)=\frac1S\sum_{j\neq l}Q_jQ_l\,\mathcal{K}(|\bX_j-\bX_l|,z),
\label{eq:pair}
\end{equation}
with the kernel
\begin{equation}
\mathcal{K}(R,z)=-\alpha_p\Big(\frac{4\pi e}{1+\varepsilon}\Big)^2\frac{z}{\pi(R^2+4z^2)^{3/2}} .
\label{eq:kernel}
\end{equation}
For disordered charges, the sum over the pairs in the area $S$ equals its average over the separations with the pair correlation function $g(R)$,
\begin{equation}
\tilde V_0^{\rm pair}(z)=n_c^2Q_c^2\!\int\!\dd^2R\,g(R)\,\mathcal{K}(R,z),
\label{eq:pairavg}
\end{equation}
up to a relative fluctuation from one configuration to another which is derived in Appendix~\ref{app:fluct},
\begin{equation}
\frac{\langle\delta\tilde V_0^2\rangle^{1/2}}{|\tilde V_0|}=\Big(\frac{4\pi z^2}{S}\Big)^{1/2} .
\label{eq:fluct}
\end{equation}
It is below $2\times10^{-2}$ at $z=5~\mu$m for a coherent area of 10~cm by 10~$\mu$m, i.e.\ the length of the mirror times the width of a collimated beam. For uncorrelated positions $g(R)=1$, the integral of the kernel does not depend on $z$, and the pair part is a constant without force; the tail (\ref{eq:C2}) is then the whole height-dependent potential. In general both parts combine into the structure factor of the charges,
\begin{equation}
S_\sigma(q)=n_cQ_c^2\Big[1+n_c\!\int\!\dd^2R\,\big(g(R)-1\big)e^{-i\bq\cdot\bm{R}}\Big],
\label{eq:Ssigma}
\end{equation}
and
\begin{equation}
\tilde V_0(z)=-\alpha_p\Big(\frac{4\pi e}{1+\varepsilon}\Big)^2\!\int\!\frac{\dd^2q}{(2\pi)^2}\,S_\sigma(q)\,e^{-2qz} .
\label{eq:V0S}
\end{equation}
The potential at the height $z$ probes the structure factor at wave numbers of the order of $1/2z$. The averaging thus requires disorder of the charges, not an ensemble of mirrors or of atoms: an atom coherent over the whole mirror sees the same tail. For charges ordered with a period $a_L$ the pair sum cancels the single-cluster part above about $a_L/10$, there is no tail, and the off-diagonal elements produce discrete diffraction orders instead.

\section{Specular reflection with the inverse-square tail}
\label{sec:specular}

\subsection{Exact solution and threshold law}

With the tail (\ref{eq:C2}), the specular equation (\ref{eq:specular}) reads
\begin{equation}
-\frac{\hbar^2}{2m}\phi''+\Big[V_{\rm CP}(z)-\frac{C_2}{z^2}\Big]\phi=\frac{\hbar^2k^2}{2m}\phi,
\label{eq:tail}
\end{equation}
with the absorbing condition (\ref{eq:absorb}). The tail enters through the dimensionless coupling
\begin{equation}
\zeta=\frac{2mC_2}{\hbar^2}=\frac{4\pi m\alpha_pe^2\,n_cQ_c^2}{\hbar^2(1+\varepsilon)^2},
\label{eq:zeta}
\end{equation}
which depends on the charges only through the product $n_cQ_c^2$ and on the isotope through the mass (Table~\ref{tab:param}). The tail exceeds the CP potential above the height
\begin{equation}
z_c=\frac{\beta_4}{\sqrt{\zeta}},
\label{eq:zc}
\end{equation}
which is 170~nm for H on Si at $10^{12}$$e$~cm$^{-2}$. Above $z_c$, Eq.~(\ref{eq:tail}) is the radial equation of a free particle with the noninteger angular momentum $l$ defined by
\begin{equation}
l(l+1)=-\zeta,\qquad \nu\equiv l+\tfrac12=\sqrt{\tfrac14-\zeta},
\label{eq:nu}
\end{equation}
and its exact solutions are
\begin{equation}
\phi(z)=c_1\sqrt z\,J_\nu(kz)+c_2\sqrt z\,J_{-\nu}(kz)\qquad (z\gg z_c).
\label{eq:bessel}
\end{equation}
Below $z_c$, if the wavelength exceeds $z_c$, the solution is the zero-energy solution with the absorbing condition, which above $z_c$ behaves as
\begin{equation}
\phi(z)\simeq z^{1/2+\nu}+B\,z^{1/2-\nu}\qquad (z_c\ll z\ll 1/k),
\label{eq:zeroE}
\end{equation}
with a complex constant $B$ fixed by the CP region; for vanishing coupling $B=-a$. Matching the small-argument form of Eq.~(\ref{eq:bessel}) to Eq.~(\ref{eq:zeroE}) gives
\begin{equation}
\varrho\equiv\frac{c_2}{c_1}=B\Big(\frac k2\Big)^{2\nu}\frac{\Gamma(1-\nu)}{\Gamma(1+\nu)},
\label{eq:rho}
\end{equation}
and the large-argument form gives the reflection coefficient
\begin{equation}
\Rr=\frac{|e^{-i\nu\pi/2}+\varrho\,e^{i\nu\pi/2}|^2}{|e^{i\nu\pi/2}+\varrho\,e^{-i\nu\pi/2}|^2},
\label{eq:Rexact}
\end{equation}
which, for a small ratio (\ref{eq:rho}), becomes
\begin{equation}
1-\Rr\simeq4\sin(\pi\nu)\frac{\Gamma(1-\nu)}{\Gamma(1+\nu)}\Big(\frac k2\Big)^{2\nu}\operatorname{Im}B .
\label{eq:fractional}
\end{equation}
This is the central result: the tail changes the linear threshold law (\ref{eq:threshold}) into the fractional power $\vperp^{2\nu}$ with the exponent
\begin{equation}
2\nu=\sqrt{1-4\zeta}=1-2\zeta+O(\zeta^2),
\label{eq:exponent}
\end{equation}
which is smaller than unity, so that the absorption at low velocity is always enhanced by the charges. For $\zeta\to0$ Eq.~(\ref{eq:fractional}) reduces to Eq.~(\ref{eq:threshold}).

\begin{figure*}[t]
\includegraphics[width=\textwidth]{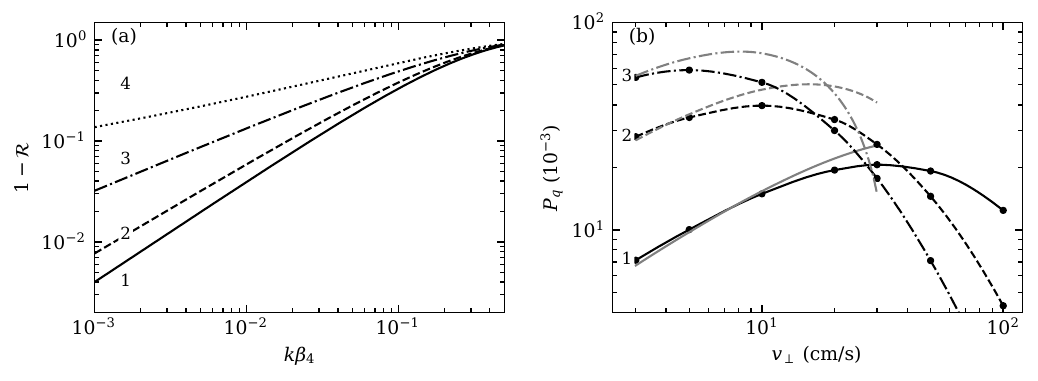}
\caption{Threshold law with the inverse-square tail. (a) Absorption $1-\Rr$ for the pure retarded potential with the tail, Eqs.~(\ref{eq:Rexact}) and (\ref{eq:rhopure}), as a function of $k\beta_4$ for the couplings $\zeta=0$ (curve 1), 0.05 (2), 0.15 (3), and 0.24 (4); the slope at small $k\beta_4$ is the exponent $2\nu$ of Eq.~(\ref{eq:exponent}). (b) Charge-induced absorption $P_q$, Eq.~(\ref{eq:Pq}), of H (1), D (2), and T (3) at the SiO$_2$/Si mirror with $10^{12}$ uncorrelated single charges per cm$^2$: exact solution of Eq.~(\ref{eq:tail}) with the Lifshitz potential (black; circles are the calculated values of Table~\ref{tab:coeff}, lines are interpolations) and logarithmic form (\ref{eq:log}) with the exact absorption of the neutral mirror (gray, shown up to 30~cm/s).}
\label{fig:threshold}
\end{figure*}

For the pure retarded potential with the tail, the zero-energy solution is
\begin{equation}
\phi(z)=\sqrt z\,H^{(1)}_\nu\Big(\frac{\beta_4}{z}\Big),
\label{eq:hankel}
\end{equation}
which gives
\begin{equation}
\varrho=-e^{-i\nu\pi}\Big[\frac{\Gamma(1-\nu)}{\Gamma(1+\nu)}\Big]^2\Big(\frac{k\beta_4}{4}\Big)^{2\nu} .
\label{eq:rhopure}
\end{equation}
Figure~\ref{fig:threshold}(a) shows the resulting absorption for several couplings. We have verified Eqs.~(\ref{eq:Rexact}) and (\ref{eq:rhopure}) against the numerical solution of Eq.~(\ref{eq:tail}) with $V_{\rm CP}=-C_4/z^4$: for $\zeta=0.02$ and $0.05$ they agree to $0.1\%$ at 3~cm/s and to $2.5\%$ at 30~cm/s, where $k\beta_4=0.1$ and the small-argument matching begins to fail. For the Lifshitz potential, we solve Eq.~(\ref{eq:tail}) numerically at each velocity (Appendix~\ref{app:numerics}), including the region where the wavelength is comparable with $z_c$ or $\beta_4$.

\subsection{Logarithmic form}
\label{sec:log}

For small coupling, Eq.~(\ref{eq:fractional}) may be expanded. The fractional power of the wave number produces a logarithm,
\begin{equation}
\Big(\frac k2\Big)^{2\nu}=\frac k2\Big[1-2\zeta\ln\frac k2+O(\zeta^2)\Big],
\label{eq:expand}
\end{equation}
and, together with the first-order change of $B$ and the pre-factor, this gives
\begin{equation}
1-\Rr=4ka''\Big[1+2\zeta\ln\frac{\mathcal{C}}{k\beta_4}\Big],
\label{eq:log}
\end{equation}
where $\mathcal{C}$ collects the contribution of the CP region. For the pure retarded potential Eq.~(\ref{eq:rhopure}) gives $\ln\mathcal{C}=2-2\gamma-2\ln2$, i.e.\ $\mathcal{C}=0.582$, with $\gamma$ Euler's constant; for the Lifshitz potentials, $\mathcal{C}$ is obtained from the exact solution at small $k$ and listed in Table~\ref{tab:param}.

The origin of the logarithm is transparent in the first-order perturbation theory. For the zero-energy solutions $u$ and $\tilde u$ of a potential and the same potential perturbed by $\delta U$, the Wronskian gives the identity
\begin{equation}
\tilde a-a=\frac{2m}{\hbar^2}\int_0^\infty u\,\delta U\,\tilde u\,\dd z ,
\label{eq:wronskian}
\end{equation}
so that, in the first order, the absorptive part is the integral of the tail with the weight $\operatorname{Im}u^2$, which, for the retarded potential, is
\begin{equation}
\operatorname{Im}u^2=z^2\sin\frac{2\beta_4}{z}\;\longrightarrow\;2\beta_4z\qquad(z\gg\beta_4).
\label{eq:weight0}
\end{equation}
The tail $-C_2/z^2$ thus contributes the integral of $1/z$, which diverges logarithmically at large heights at zero energy. At finite $k$ the weight becomes
\begin{equation}
\operatorname{Im}\chi^2=\frac{a''\sin2kz}{k}-\frac{2a'\sin^2kz}{k},
\label{eq:weightk}
\end{equation}
and
\begin{equation}
\frac{a''}{k}\int_{z_0}^\infty\!\frac{\sin2kz}{z^2}\,\dd z
=2a''\Big[\ln\frac{1}{2kz_0}+1-\gamma\Big]+O(kz_0),
\label{eq:logint}
\end{equation}
with $z_0$ between $\beta_4$ and $1/k$. The energy thus cuts the logarithm at the height
\begin{equation}
z_{\max}=\frac{c_\gamma}{k},\qquad c_\gamma=\frac{e^{1-\gamma}}{2}=0.763 ,
\label{eq:zmax}
\end{equation}
and the lower cutoff, of the order of $\beta_4$, is contained in $\mathcal{C}$. The charge effect is therefore accumulated between $\beta_4$ and $z_{\max}$, i.e.\ over one to two decades of height, which is why it is much larger than the ratio of the tail to the CP potential at the reflection height would suggest.

Equation (\ref{eq:log}) requires small values of both $\zeta\ln(\mathcal{C}/k\beta_4)$ and $k\beta_4$. Figure~\ref{fig:threshold}(b) compares it with the exact solution for the Lifshitz potential. At 3~cm/s, it is accurate to 6\% for H and D and to 11\% for T; at 10~cm/s and above, where $k\beta_4$ exceeds 0.03 (H), 0.1 (D), and 0.2 (T), it overestimates the charge effect substantially, and the exact solution is required.

\subsection{Absorption coefficients}
\label{sec:coeff}

\begin{table*}[t]
\caption{ Absorption coefficient $1-\Rr$, in units of $10^{-3}$, of H, D, and T at the SiO$_2$(2~nm)/Si mirror ($\varepsilon=11.7$, upper block) and at the MgF$_2$ mirror ($\varepsilon=5.2$, lower block) for uncorrelated single charges of density $N$ (per cm$^2$), exact solution of Eq.~(\ref{eq:tail}) with the Lifshitz potentials; the columns $N=0$ refer to the neutral mirror. The results hold for longitudinal velocities of 100--300~m/s. For clusters of $Q_c$ charges the density $N$ is to be multiplied by $Q_c$.}
\label{tab:coeff}
\begin{ruledtabular}
\begin{tabular}{ddddddddddddd}
\multicolumn{1}{c}{$\vperp$} & \multicolumn{4}{c}{H, $N$} & \multicolumn{4}{c}{D, $N$} & \multicolumn{4}{c}{T, $N$} \\
\multicolumn{1}{c}{(cm/s)} & \multicolumn{1}{c}{0} & \multicolumn{1}{c}{$10^{10}$} & \multicolumn{1}{c}{$10^{11}$} & \multicolumn{1}{c}{$10^{12}$} & \multicolumn{1}{c}{0} & \multicolumn{1}{c}{$10^{10}$} & \multicolumn{1}{c}{$10^{11}$} & \multicolumn{1}{c}{$10^{12}$} & \multicolumn{1}{c}{0} & \multicolumn{1}{c}{$10^{10}$} & \multicolumn{1}{c}{$10^{11}$} & \multicolumn{1}{c}{$10^{12}$} \\
\hline
\multicolumn{13}{l}{SiO$_2$(2 nm)/Si} \\
3   & 43.06 & 43.13 & 43.72 & 50.19 & 119.3 & 119.6 & 121.9 & 147.4 & 207.5 & 208.0 & 212.5 & 261.8 \\
5   & 70.67 & 70.76 & 71.61 & 80.69 & 190.0 & 190.3 & 193.2 & 224.8 & 317.9 & 318.5 & 323.5 & 376.8 \\
10  & 135.9 & 136.0 & 137.3 & 150.8 & 339.0 & 339.3 & 342.8 & 378.7 & 520.5 & 521.1 & 525.7 & 571.9 \\
20  & 250.9 & 251.1 & 252.8 & 270.3 & 546.6 & 547.0 & 550.0 & 580.6 & 739.9 & 740.2 & 743.0 & 770.0 \\
30  & 347.7 & 347.9 & 349.7 & 368.3 & 676.9 & 677.1 & 679.5 & 702.7 & 844.2 & 844.4 & 846.1 & 861.9 \\
50  & 497.9 & 498.0 & 499.8 & 517.1 & 819.6 & 819.7 & 821.0 & 834.1 & 932.1 & 932.2 & 932.9 & 939.2 \\
100 & 716.0 & 716.2 & 717.3 & 728.4 & 940.1 & 940.1 & 940.5 & 944.4 & 984.5 & 984.5 & 984.7 & 985.9 \\
\hline
\multicolumn{13}{l}{MgF$_2$} \\
3   & 25.47 & 25.66 & 27.44 & 54.29 & 73.33 & 74.14 & 81.77 & 213.6 & 131.6 & 133.3 & 149.6 & 422.9 \\
5   & 42.07 & 42.35 & 44.95 & 81.93 & 118.9 & 120.0 & 130.3 & 284.6 & 208.3 & 210.5 & 230.5 & 504.9 \\
10  & 82.23 & 82.68 & 86.84 & 141.2 & 221.9 & 223.4 & 237.3 & 409.9 & 366.9 & 369.4 & 391.8 & 632.2 \\
20  & 156.9 & 157.6 & 163.7 & 237.0 & 387.1 & 388.8 & 404.1 & 565.2 & 579.9 & 582.0 & 600.4 & 765.6 \\
30  & 224.5 & 225.3 & 232.5 & 314.4 & 510.1 & 511.7 & 525.7 & 662.5 & 708.0 & 709.6 & 723.1 & 836.8 \\
50  & 340.6 & 341.5 & 349.5 & 435.0 & 673.3 & 674.5 & 684.9 & 778.9 & 842.7 & 843.6 & 850.8 & 909.3 \\
100 & 546.1 & 546.9 & 554.1 & 625.0 & 856.0 & 856.5 & 861.1 & 900.9 & 950.4 & 950.6 & 952.7 & 969.1 \\
\end{tabular}
\end{ruledtabular}
\end{table*}

Table~\ref{tab:coeff} gives the absorption coefficient of the three isotopes at the two mirrors for charge densities from $10^{10}$ to $10^{12}$~$e/$cm$^2$ and normal velocities of 3--100~cm/s, and Fig.~\ref{fig:absorption} shows the neutral and the charged mirrors together. We define the charge-induced absorption
\begin{equation}
P_q=\Rr_{\rm CP}-\Rr,
\label{eq:Pq}
\end{equation}
with $\Rr_{\rm CP}$ the reflection coefficient of the neutral mirror, and, in the linear regime $\zeta\ln(\mathcal{C}/k\beta_4)\ll1$, the reduced charge-induced absorption $y_q$ by
\begin{equation}
P_q=4k\,\frac{N}{10^{10}\,{\rm cm}^{-2}}\;y_q(\vperp),
\label{eq:yq}
\end{equation}
which has dimension of length and plays the role of the charge-induced increase of $a''$~$10^{10}$ $e/$cm$^2$.

The charge effect is the largest at lower velocities and for heavier isotopes. On Si at 3~cm/s and $10^{12}$~$e/$cm$^{2}$, it increases absorption by 17\% (H), 24\% (D), and 26\% (T); at $10^{11}$~$e/$cm$^{2}$ by 1.5\%, 2.2\%, and 2.4\%. It decreases with velocity, because the reflection moves into the CP region, where the tail is weak, and because the cutoff $z_{\max}$ of Eq.~(\ref{eq:zmax}) approaches $\beta_4$. On MgF$_2$, the effect is much stronger for two reasons: the tail is 4.2 times stronger at the same density because of the smaller permittivity [Eq.~(\ref{eq:zeta})], and the intrinsic absorption is smaller. At 3~cm/s, the charges increase the absorption by 8\%, 12\%, and 14\% at $10^{11}$~$e/$cm$^{2}$ and, at $10^{12}$~$e/$cm$^{2}$, double it for H and triple it for D and T. The coupling then reaches 0.076, 0.151, and 0.226 (Table~\ref{tab:param}), so that T on MgF$_2$ approaches the critical value $\zeta=1/4$ of Sec.~\ref{sec:strong}. The isotope effect and the correlations are discussed below for Si; for MgF$_2$ they are qualitatively the same.

\subsection{Isotope effect}
\label{sec:isotope}

\begin{figure}[t]
\includegraphics[width=\columnwidth]{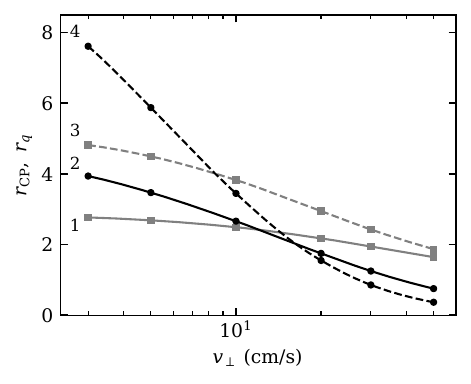}
\caption{Isotope ratios (\ref{eq:ratios}) on the SiO$_2$/Si mirror as a function of the normal velocity: intrinsic ratio $r_{\rm CP}$ (gray, squares) and charge-induced ratio $r_q$ at $10^{12}$ charges per cm$^2$ (black, circles) for deuterium (curves 1 and 2, solid) and tritium (curves 3 and 4, dashed). Symbols are the values from Table~\ref{tab:coeff}; lines interpolate. The charge-induced ratio exceeds the intrinsic one below 13~cm/s (D) and 9~cm/s (T).}
\label{fig:isotope}
\end{figure}

We compare the isotope X with H at the same normal velocity through the ratios of the intrinsic and of the charge-induced absorption,
\begin{equation}
r^{\rm X}_{\rm CP}=\frac{1-\Rr_{{\rm CP},{\rm X}}}{1-\Rr_{{\rm CP},{\rm H}}},\qquad
r^{\rm X}_q=\frac{P_{q,{\rm X}}}{P_{q,{\rm H}}} .
\label{eq:ratios}
\end{equation}
At threshold, Eq.~(\ref{eq:threshold}) gives the intrinsic ratio
\begin{equation}
r^{\rm X}_{\rm CP}=\frac{m_{\rm X}a''_{\rm X}}{m_{\rm H}a''_{\rm H}},
\label{eq:rCP}
\end{equation}
2.90 for D and 5.33 for T. The charge-induced ratio is larger, 3.9 and 7.6, at 3~cm/s, because the coupling (\ref{eq:zeta}) contains an additional factor of mass, so that, in the linear regime, $P_q$ scales roughly as $m^2a''$. The two ratios cross at 13~cm/s (D) and 9~cm/s (T) (Fig.~\ref{fig:isotope}). A heavier isotope thus enhances the charge effect relative to the intrinsic absorption at low velocity, and a comparison of two isotopes at the same normal velocity separates the charge contribution from the CP one.

\subsection{Correlated charges}
\label{sec:corr}

\begin{figure*}[t]
\includegraphics[width=\textwidth]{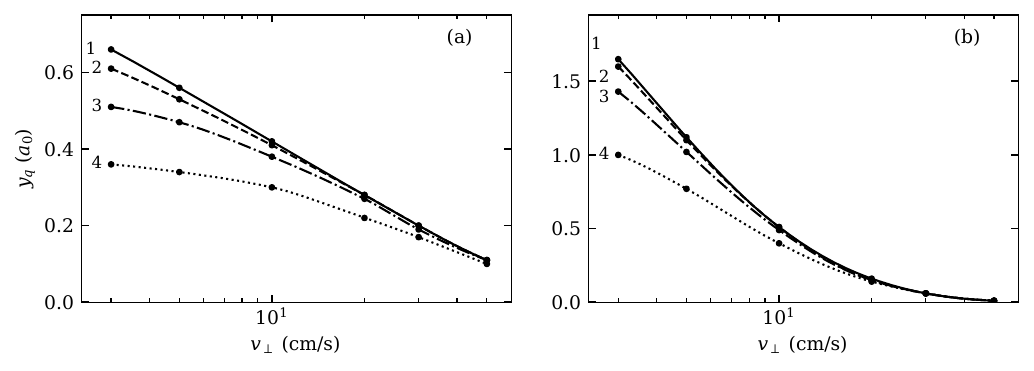}
\caption{ Reduced charge-induced absorption $y_q$, Eq.~(\ref{eq:yq}), on the SiO$_2$/Si mirror in the linear regime for (a) H and (b) T, for charges neutralized over the length $\xi$ with the structure factor (\ref{eq:Sxi}): uncorrelated charges, $\xi\to\infty$ (curves 1), and $\xi=3~\mu$m (2), $1~\mu$m (3), and $0.3~\mu$m (4). Symbols: calculated values; lines: interpolation. The correlations cut the logarithm of Eq.~(\ref{eq:log}) at the height $\xi$ and saturate the effect at low velocity.}
\label{fig:corr}
\end{figure*}

For charges neutralized on the length $\xi$, e.g.\ by compensating charges of the opposite sign, we take the structure factor
\begin{equation}
S_\sigma(q)=n_cQ_c^2\,\frac{q^2\xi^2}{1+q^2\xi^2},
\label{eq:Sxi}
\end{equation}
which vanishes at small wave numbers, as it must for a neutral system, and tends to the uncorrelated value at $q\xi\gg1$. Equation (\ref{eq:V0S}) then gives
\begin{equation}
\tilde V_0(z)=-\frac{C_2}{z^2}\,f\Big(\frac z\xi\Big),\qquad
f(s)=4s^2\!\int_0^\infty\!\frac{t^3}{1+t^2}\,e^{-2ts}\,\dd t,
\label{eq:fcorr}
\end{equation}
with the limits
\begin{equation}
f(s)\to1\quad(s\ll1),\qquad f(s)\simeq\frac{3}{2s^2}\quad(s\gg1).
\label{eq:flimits}
\end{equation}
Beyond $\xi$, the tail falls as the inverse fourth power of height, and the logarithm in Eq.~(\ref{eq:log}) is cut at the smaller of $z_{\max}$ and $\xi$. The reduced absorption therefore saturates below the velocity at which $z_{\max}$ reaches $\xi$,
\begin{equation}
v_{\rm sat}\simeq c_\gamma\,\frac{\hbar}{m\xi},
\label{eq:vsat}
\end{equation}
which is 4.8~cm/s for H at $\xi=1~\mu$m. Figure~\ref{fig:corr} shows $y_q$ for correlation lengths of 3, 1, and 0.3~$\mu$m: at 3~cm/s, a correlation length of $1~\mu$m reduces the charge effect by 23\% (H) and 13\% (T), and $0.3~\mu$m by 45\% and 39\%, whereas above 30~cm/s the correlations change $y_q$ by less than 15\% even for $\xi=0.3~\mu$m. The velocity dependence of the charge effect at low velocity thus carries information on the compensation length of the charges.

\subsection{Strong coupling}
\label{sec:strong}

Above the critical coupling $\zeta=1/4$, the order (\ref{eq:nu}) becomes imaginary,
\begin{equation}
\nu=i\mu,\qquad \mu=\sqrt{\zeta-\tfrac14},
\label{eq:mu}
\end{equation}
which is the well-known ``fall to the center'' of the inverse-square potential \cite{LandauQM,Case1950,Frank1971}. The two solutions (\ref{eq:bessel}) are then waves running towards and away from the surface in the tail, and Eq.~(\ref{eq:Rexact}) gives a reflection coefficient that oscillates log-periodically with the velocity, with the period
\begin{equation}
\Delta\ln\vperp=\frac{\pi}{\mu},
\label{eq:period}
\end{equation}
around the energy-independent value
\begin{equation}
\Rr_\infty=e^{-2\pi\mu}
\label{eq:Rinf}
\end{equation}
that an ideally absorbing inner region would give. For single charges on Si, this regime requires more than $4.6\times10^{12}$~$e/$cm$^2$ even for $T$, and $1.1\times10^{12}$~$e/$$cm^2$ on MgF$_2$; for clusters, the required density is smaller by $Q_c$. It is not reached in the beam experiments, but it marks the point where the quantum reflection of the heavier isotopes from a strongly charged MgF$_2$ mirror ceases to obey a threshold law.

\section{Diffuse scattering}
\label{sec:diffuse}

The off-diagonal elements of Eq.~(\ref{eq:channels}) carry flux into the channels open for negative $q_x$, whose normal wave numbers are
\begin{equation}
k_{\bq}=\Big(\frac{2m\vpar|q_x|}{\hbar}\Big)^{1/2} .
\label{eq:kq}
\end{equation}
With the Green function of an open channel,
\begin{equation}
G_{\bq}(z,z')=\frac{2m}{\hbar^2}\,\chi_{k_q}(z_<)\,e^{ik_qz_>},
\label{eq:green}
\end{equation}
where $\chi_{k_q}$ satisfies the absorbing condition and $z_<$, $z_>$ are the smaller and the larger of the two heights, the first-order amplitude into channel $\bq$ is
\begin{equation}
M_{\bq}=\frac{2m}{\hbar^2}\int_0^\infty\!\chi_{k_q}(z)\,\tilde V_{\bq}(z)\,\chi_k(z)\,\dd z,
\label{eq:Mq}
\end{equation}
and the diffuse loss is
\begin{equation}
P_{\rm diff}=4k\sum_{q_x<0}k_{\bq}\,|M_{\bq}|^2 .
\label{eq:Pdiff}
\end{equation}
For the single-cluster terms in Eq.~(\ref{eq:Vq}) and disordered clusters, the cross terms between different clusters average out, and
\begin{equation}
P_{\rm diff}=4k\,n_c\!\int_{q_x<0}\!\frac{\dd^2q}{(2\pi)^2}\,k_{\bq}\,|m_{\bq}|^2,
\label{eq:Pdiffavg}
\end{equation}
where $m_{\bq}$ is the matrix element (\ref{eq:Mq}) with the Fourier transform of the potential (\ref{eq:vcluster}) of one cluster; for correlated positions, $n_c$ is replaced by the structure factor of the cluster positions at the wave vector $\bq$. For one frozen configuration, the intensities of individual channels form a speckle pattern, while the total loss (\ref{eq:Pdiffavg}) does not depend on the configuration.

The specular and diffuse channels thus probe the charges differently. Specular absorption depends on the product $n_cQ_c^2$ only [Eq.~(\ref{eq:zeta})], while Eq.~(\ref{eq:Pdiffavg}) is proportional to $n_cQ_c^4$; their ratio measures the charge of the cluster. The diffuse channels have large wave numbers (\ref{eq:kq4}) and follow the phase of the threshold wave only below about 3~nm, where the cluster potential is weak, so the diffuse loss is small. Table~\ref{tab:diffuse} gives its ratio to the specular charge-induced absorption for clusters of $Q_c=32$ charges, computed from the full channel equations (\ref{eq:channels}) for a model CP potential with the same $C_3$ and $C_4$ and rescaled to the Lifshitz potential by the ratio of the first-order values for the two potentials, 0.91 (H) and 0.94 (D). For single charges, the ratio is $10^{3}$ times smaller. The diffusely scattered atoms leave at elevation angles of $13^\circ$ and $8^\circ$ at 100 and 300~m/s.

A net charge density $N_{\rm net}$ creates the uniform field
\begin{equation}
\langle E_z\rangle=\frac{4\pi eN_{\rm net}}{1+\varepsilon},
\label{eq:Enet}
\end{equation}
and the polarization energy (\ref{eq:Vch}) contains, besides a constant, the cross term
\begin{equation}
V_\times(\br,z)=-\alpha_p\langle\bE\rangle\cdot\delta\bE(\br,z),
\label{eq:cross}
\end{equation}
linear in the field fluctuation, with the Fourier components
\begin{equation}
\tilde v_\times(\bq,z)=-\alpha_p\langle E_z\rangle\,\frac{4\pi e}{1+\varepsilon}\,\frac{\sigma_{\bq}}{S}\,e^{-qz} .
\label{eq:crossq}
\end{equation}
Its average vanishes, so it does not enter the specular channel; its diffuse loss is proportional to the density cube for a fully uncompensated surface. By detuning (\ref{eq:detuning}), components with wave numbers of the order of the inverse height couple to hard channels, and those with small detuning are nearly uniform in height and only deflect the atom in the mirror plane. Counting as losses the transfers that at least double the normal velocity, Table~\ref{tab:diffuse} gives this loss for a fully uncompensated density of $10^{12}$~$e/$cm$^2$. In all cases, the diffuse losses are below $10^{-3}$ of the specular charge effect.

\begin{table}[t]
\caption{ Diffuse losses relative to the charge-induced specular absorption $P_q$ on Si. Left: single-cluster term (\ref{eq:Pdiffavg}) for clusters of $Q_c=32$ charges at the normal velocity of 1~cm/s (for single charges the values are $10^{3}$ times smaller). Right: cross term (\ref{eq:cross}) for a fully uncompensated density of $10^{12}$~$e/$cm$^2$ at the longitudinal velocity of 200~m/s.}
\label{tab:diffuse}
\begin{ruledtabular}
\begin{tabular}{lcccc}
 & \multicolumn{2}{c}{$P_{\rm diff}/P_q$} & \multicolumn{2}{c}{$P_\times/P_q$} \\
 & 100 m/s & 300 m/s & 1 cm/s & 10 cm/s \\
\hline
H & $2.9\times10^{-4}$ & $1.1\times10^{-4}$ & $1.9\times10^{-4}$ & $3.7\times10^{-4}$ \\
D & $9.4\times10^{-5}$ & $3.4\times10^{-5}$ & $3.7\times10^{-4}$ & $1.1\times10^{-3}$ \\
\end{tabular}
\end{ruledtabular}
\end{table}

\section{Conclusion}
\label{sec:concl}

From the exact coupled-channel equations, we derived how surface charges that are not compensated locally act on the quantum reflection of hydrogen-isotope beams at grazing incidence. For beams moving along the mirror at 100--300~m/s, the laterally scattered channels are detuned from the specular one by two to three orders of magnitude more than the energy resolution of the reflection region, which is set by the phase time $\tau_r$. The specular wave therefore feels only the laterally averaged polarization potential, which, for disordered charges, is the inverse-square tail $-C_2/z^2$; the tail follows exactly from the potential of single charges summed over their impact parameters, and correlations modify it through the charge structure factor.

The specular problem is then the radial problem with a non-integer angular momentum. Its exact solution replaces the linear threshold law of the neutral mirror with a fractional power of the normal velocity with the exponent $2\nu=(1-4\zeta)^{1/2}$, where the coupling $\zeta$ is proportional to the charge density, square of the cluster charge and atomic mass. The expansion of this law yields a logarithmic velocity dependence of the charge-induced absorption, which holds at the lower end of the velocity range only; Table~\ref{tab:coeff} gives the exact absorption coefficients. Above $\zeta=1/4$, the threshold law is lost and the reflection coefficient oscillates log-periodically with velocity.

Numerically, the charges increase the absorption most at low velocities and for heavier isotopes: by up to a quarter at $10^{12}$ charges per cm$^2$ on silicon, and by a factor of two to three on MgF$_2$, where the smaller permittivity strengthens the tail and the intrinsic absorption is smaller. Charges neutralized over a length $\xi$ saturate the effect below velocity (\ref{eq:vsat}), and diffuse scattering removes less than $10^{-3}$ of charge-induced absorption.

Three consequences follow for the experiments. First, a mirror with $10^{11}$ charges per cm$^2$, typical of a good thermal oxide, changes the absorption of hydrogen by 1--2\%, which is below the present accuracy of the beam experiments \cite{Killian2024}, 
whereas $10^{12}$~cm$^{-2}$ on silicon or $10^{11}$~cm$^{-2}$ on MgF$_2$ produces a change of 10--20\% that would be visible as an excess absorption at the lowest normal velocities. Second, the charge contribution has a distinct signature: its velocity dependence is a fractional power, or a logarithm, rather than the linear threshold law, and it grows with the isotope mass faster than the intrinsic absorption, so that measurements with two isotopes at the same normal velocity separate the two contributions. Third, since specular absorption depends on $n_cQ_c^2$ and diffuse loss on $n_cQ_c^4$, a measurement of both would determine the cluster charge. For gravitational quantum states, where the normal velocities are of the order of centimeters per second, the same tail acts, and the results of this paper give its effect on the state lifetimes once the charge density of the mirror is known.

\appendix

\section{Lifshitz potential }
\label{app:lifshitz}

The CP potential of a ground-state H atom above a dielectric half-space is \cite{Lifshitz1956,DLP1961}
\begin{align}
V_{\rm CP}(z)=-\frac{\hbar}{2\pi c^3}\int_0^\infty\!\dd\xi\,\xi^3\alpha(i\xi)\int_1^\infty\!\dd p\,e^{-2p\xi z/c}\nonumber\\
\times\big[(2p^2-1)r_{\rm TM}-r_{\rm TE}\big],
\label{eq:lifshitz}
\end{align}
with $c$ the speed of light and the reflection coefficients
\begin{equation}
r_{\rm TM}=\frac{\varepsilon p-s}{\varepsilon p+s},\qquad
r_{\rm TE}=\frac{p-s}{p+s},\qquad
s=\sqrt{p^2+\varepsilon-1},
\label{eq:fresnel}
\end{equation}
where $\varepsilon$ is taken at the imaginary frequency. The polarizability and the dielectric function of silicon are represented by one oscillator each,
\begin{equation}
\alpha(i\xi)=\frac{\alpha_p}{1+(\xi/\omega_a)^2},\qquad
\varepsilon(i\xi)=1+\frac{\varepsilon_0-1}{1+(\xi/\omega_{\rm UV})^2},
\label{eq:oscillators}
\end{equation}
with $\hbar\omega_a=12.09$~eV, $\varepsilon_0=11.87$, and $\hbar\omega_{\rm UV}=4.35$~eV, the latter fixed by the value of $C_3$. The value $\varepsilon_0=11.87$ is the one that reproduces the measured van der Waals coefficient with a single oscillator; it differs by 1.5\% from the static permittivity 11.7 used for the field of the charges, which changes $\zeta$ by 3\%, well within the uncertainty of the charge density. The 2~nm oxide changes the potential only below 2~nm and is neglected. For MgF$_2$, the infrared response matters, and the dielectric function contains an infrared term with a strength of 3.3 at 0.052~eV and an ultraviolet term with a strength of 0.9 at 14.6~eV, the latter fixed by the van der Waals coefficient $C_3=0.0466$; the static and optical permittivities are 5.2 and 1.9. A two-oscillator representation of the dielectric functions changes the imaginary scattering lengths by less than 1\% and the real parts by up to 1~nm.

\section{Fluctuation of the averaged potential }
\label{app:fluct}

For uncorrelated positions, the Fourier components (\ref{eq:sigma}) of the charge density are, for $q\neq0$, sums of $NS$ unit phases, so for $NS\gg1$ they are Gaussian with $\langle\sigma_{\bq}\sigma^*_{\bq'}\rangle=NS\,\delta_{\bq\bq'}$ and $\sigma_{-\bq}=\sigma^*_{\bq}$. The variance of $|\sigma_{\bq}|^2$ is then $(NS)^2$, and the pair $\pm\bq$ is fully correlated. Inserting this into Eq.~(\ref{eq:V0}) and converting the sums with Eq.~(\ref{eq:fourier}) gives
\begin{align}
\langle\delta\tilde V_0^2\rangle&=2\alpha_p^2\Big(\frac{4\pi e}{1+\varepsilon}\Big)^4\frac{N^2}{S}\int\!\frac{\dd^2q}{(2\pi)^2}\,e^{-4qz}\nonumber\\
&=\frac{\alpha_p^2}{16\pi}\Big(\frac{4\pi e}{1+\varepsilon}\Big)^4\frac{N^2}{Sz^2},
\label{eq:varV0}
\end{align}
while the mean is $\tilde V_0=-\alpha_p[4\pi e/(1+\varepsilon)]^2N/(8\pi z^2)$. The ratio of the standard deviation to the mean is Eq.~(\ref{eq:fluct}). The fluctuation increases with height because the potential at height $z$ is formed by charges within an area of the order of $z^2$, whose number is $Nz^2$, while the coherence area contains $NS$ charges.

\section{Numerical methods}
\label{app:numerics}

The Lifshitz potential is tabulated on a logarithmic grid from 2 to $3\times10^6\,a_0$ and interpolated by a cubic spline. The absorbing condition (\ref{eq:absorb}) is imposed at height $z_{\rm in}=6\,a_0$; Eq.~(\ref{eq:tail}) is integrated by the log-derivative method \cite{Johnson1973,Manolopoulos1986} to $4000\,a_0$ and by an adaptive integrator to $300/k$, where the solution is matched to incoming and outgoing waves. Moving $z_{\rm in}$ between 4 and $12\,a_0$ changes $a''$ by less than 0.4\% (0.1~nm for H), which sets the numerical accuracy of the tabulated absorption coefficients. The channel equations (\ref{eq:channels}) for single clusters were solved by direct time propagation along the classical trajectory and as coupled channels, with identical results.

\end{document}